\documentclass[conference]{IEEEtran}
\IEEEoverridecommandlockouts

\usepackage{algorithmic}
\usepackage{algorithm}
\usepackage{libertine}
\usepackage[libertine]{newtxmath}
\usepackage{amsmath, amsfonts, mathtools}
\usepackage{array}
\usepackage[font=bf, labelfont=bf, textfont=bf]{caption}
\usepackage[sort, nocompress]{cite}
\usepackage[scaled=0.965]{couriers}
\usepackage{enumitem}
\usepackage{fancyhdr}
\usepackage{fontenc}
\usepackage[shortcuts, acronym]{glossaries}
\usepackage{graphicx}
\usepackage[innertopmargin=0.12cm, innerbottommargin=0.15cm, innerleftmargin=0.12cm, innerrightmargin=0.12cm]{mdframed}
\usepackage{multirow}
\usepackage{soul}
\usepackage{textcomp}
\usepackage[normalem]{ulem}
\usepackage{verbatim}
\usepackage{xcolor}
\usepackage{threeparttable}

\usepackage{subcaption}

\def\BibTeX{{\rm B\kern-.05em{\sc i\kern-.025em b}\kern-.08em
    T\kern-.1667em\lower.7ex\hbox{E}\kern-.125emX}}

\DeclareCaptionFont{captionfont}{\fontsize{9pt}{10pt}\selectfont}
\renewcommand{\footnoterule}{
  \kern -3pt
  \hrule width 3.5in height 0.3pt
  \kern 3pt
}

\begin{document}

\title{TriGen: NPU Architecture for End-to-End Acceleration of Large Language Models based on SW-HW Co-Design}
\author{
    \IEEEauthorblockN{Jonghun Lee\IEEEauthorrefmark{1}\thanks{This work was primarily conducted while all authors were affiliated with Samsung Research.}, Junghoon Lee\IEEEauthorrefmark{2}, Hyeonjin Kim\IEEEauthorrefmark{3}, Seoho Jeon\IEEEauthorrefmark{3}, Jisup Yoon\IEEEauthorrefmark{2}, Hyunbin Park\IEEEauthorrefmark{3},\\
    Meejeong Park\IEEEauthorrefmark{2}, and Heonjae Ha\IEEEauthorrefmark{2}}
    \IEEEauthorblockA{\IEEEauthorrefmark{1}Gachon University, Seongnam, Korea, jonghun@gachon.ac.kr}
    \IEEEauthorblockA{\IEEEauthorrefmark{2}Samsung Electronics, Suwon, Korea, \{jayhoon.lee, jisup.yoon, heonjae.ha\}@samsung.com}
    \IEEEauthorblockA{\IEEEauthorrefmark{3}Samsung Research, Seoul, Korea, \{hyeonjin.kim, seoho.jeon, h\_bin.park, mia.park\}@samsung.com}
}
\maketitle

\thispagestyle{plain}
\pagestyle{plain}

\newcommand{\iscasubmissionnumber}{NaN}
\author{\normalsize{ISCA 2026 Submission
    \textbf{\#\iscasubmissionnumber} -- Confidential Draft -- Do NOT Distribute!!}}

\hyphenation{TriGen}
\hyphenation{Figure}
\hyphenation{Table}

\begin{abstract}
Recent studies have extensively explored \ac{NPU} architectures for accelerating AI inference in on-device environments, which are inherently resource-constrained. 
Meanwhile, transformer-based large language models (LLMs) have become dominant, with rapidly increasing model sizes but low degree of parameter reuse compared to conventional CNNs, making end-to-end execution on resource-limited devices extremely challenging.
To address these challenges, we propose \textit{TriGen}, a novel NPU architecture tailored for resource-constrained environments through software-hardware co-design.
Firstly, TriGen adopts low-precision computation using microscaling (MX) to enable additional optimization opportunities while preserving accuracy, and resolves the issues that arise by employing such precision.
Secondly, to jointly optimize both nonlinear and linear operations, TriGen eliminates the need for specialized hardware for essential nonlinear operations by using fast and accurate \ac{LUT}, thereby maximizing performance gains and reducing hardware-cost in on-device environments, and finally, by taking practical hardware constraints into account, further employs scheduling techniques to maximize computational utilization even under limited on-chip memory capacity.
We evaluate the performance of TriGen on various LLMs and show that TriGen achieves an average 2.73$\times$ performance speedup and 52\% less memory transfer over the baseline NPU design with negligible accuracy loss.

\end{abstract}


\newacronym{ACC}{ACC}{accumulator}
\newacronym{CP}{CP}{control processor}
\newacronym{CWQ}{CWQ}{channel-wise quantization}
\newacronym{DLA}{DLA}{deep learning accelerator}
\newacronym{DNN}{DNN}{deep neural network}
\newacronym{DSP}{DSP}{digital signal processor}
\newacronym{FC}{FC}{fully-connected}
\newacronym{FFN}{FFN}{feed forward network}
\newacronym{FI32}{FI32}{Floating Integer 32-bit}
\newacronym{GEMM}{GEMM}{general matrix multiplication}
\newacronym{INV}{INV}{Inverse}
\newacronym{ISA}{ISA}{Instruction Set Architecture}
\newacronym{LLM}{LLM}{large language model}
\newacronym{LUT}{LUT}{lookup table}
\newacronym{MAC}{MAC}{multiply and accumulate}
\newacronym{MLP}{MLP}{multi-layer perceptron}
\newacronym{MPA}{MPA}{MAC processing array}
\newacronym{MX}{MX}{microscaling}
\newacronym{NPU}{NPU}{neural processing unit}
\newacronym{PSUM}{PSUM}{Partial Summation}
\newacronym{RMSNorm}{RMSNorm}{root mean square normalization}
\newacronym{RoPE}{RoPE}{rotary position embeddings}
\newacronym{SFU}{SFU}{special function unit}
\newacronym{VPU}{VPU}{vector processing unit}
\newacronym{SiLU}{SiLU}{sigmoid linear unit}
\newacronym{SR}{SR}{square root}
\newacronym{PE}{PE}{processing element}
\newacronym{PPA}{PPA}{post-processing array}
\newacronym{PPL}{PPL}{Perplexity}
\newacronym{TMU}{TMU}{tensor manipulation unit}
\newacronym{TOPS}{TOPS}{tera operations per second}
\newacronym{MAPE}{MAPE}{Mean Absolute Percentage Error}
\newacronym{MSE}{MSE}{Mean Squared Error}
\newacronym{matmul}{matmul}{matrix multiplication}
\newacronym{SDPA}{SDPA}{Scaled Dot-Product Attention}

\section{Introduction} \label{sec:introduction}
In recent years, numerous research efforts have been conducted to develop accelerators for \acp{LLM}.
Building upon this, numerous quantization studies have been carried out to reduce the data precision used in \ac{LLM} accelerators.
Although various studies have been conducted on different combinations of data precision, most of these studies have been carried out without seriously considering both computational accuracy and given resource constraints simultaneously.

For instance, in most studies, while weight parameters are quantized to 4-bit or lower (e.g., weights-only quantization), activations are maintained in FP16 to preserve accuracy \cite{xu_asplos2025, jang_hpca2024, xilong_arxiv2025, fang_hpca2025, dettmers_icml2023, frantar_iclr2023, lin_mlsys2024, lin_mlsys2025}, resulting in no reduction in memory consumption for activations.
Even when activations are quantized to 8-bit and weights to 4-bit or lower, reducing memory consumption, the outlier portions of activations are still typically processed in 16-bit to ensure accuracy \cite{dettmers_neurips2022, liu_iclr2025, xiao_icml2023, park_isca2018, lee_date2025, guo_isca2023, li_tcas2024, ramachandran_isca2025}.  
In the same vein, it is common practice in actual commercial products to still use FP16 activations.
While the use of FP16 for \ac{LLM} computations typically leads to longer latency and higher power consumption due to the increased computation and memory usage, this issue is significantly more pronounced in resource-limited systems (e.g., on-device environments).
Therefore, reducing the precision of both activations and weights while maintaining accuracy is essential and critical.

Reducing the precision of data from 16-bit to 8-bit theoretically halves the computational load and memory usage, and it offers the advantage of reducing the total amount of data used in the KV cache.
Especially in on-device environments where SRAM is highly limited, reducing the precision of data provides the compiler with more exploration options, leading to overall performance improvements in \ac{LLM} accelerators.
From this perspective, the adoption of \textit{\ac{MX}} \cite{darvish2023shared, rouhani_arxiv2023, rouhani2023ocp} can emerge as a highly effective and strategic choice to maintain accuracy while reducing memory consumption.
Moreover, addressing the various challenges that arise when introducing \acp{MX} in constrained environments, such as on-device systems, is of critical importance.

Secondly, in a typical transformer-based \ac{LLM} models, linear operations such as \ac{matmul} account for the majority of computations, yet they are executed with high efficiency due to the utilization of extensive MAC arrays and the inherent simplicity of the operations themselves, with consuming minimal latency.
Conversely, nonlinear operations such as softmax, \ac{SiLU} \cite{elfwing_nn2018}, and normalization \cite{ba_arXiv2016, zhang_neurips2019} in \acp{LLM} involve higher computational complexity and are often required to be processed by separate processors or dedicated hardware units rather than directly within the LLM accelerator.
As a result, the proportion of latency consumed by these operations tends to be significant.
If the nonlinear operations could be processed within a single unit rather than relying on separate processors, it would significantly contribute to enhancing overall performance.
The seamless and appropriate utilization of \acp{LUT} within LLM accelerators could emerge as an initial step \cite{Xie2020Twofold, yu_dac2022, kim_dac2023}.

The approach of processing nonlinear operations using \acp{LUT} can eliminate the overhead associated with data precision expansion and conversion.
Additionally, it can prevent the inherent data movement that is unavoidable in certain architectural designs.
To achieve this, it is essential to design \ac{LUT} that is not only simple but also capable of delivering high-accuracy results.
Furthermore, it is crucial to approach architectural design based on a software-level understanding of the nonlinear operations employed in the model.

Lastly, while existing NPUs often pursue peak TOPS, their effective performance on memory-limited devices is largely constrained by how data are moved and reused.
Under tight on-chip SRAM and limited DRAM bandwidth, poor resource allocation—such as sub-optimal tile sizes or mis-chosen stationary policies—can negate the advantage of high compute density, leading to pipeline stalls and excessive DRAM access.
In on-device LLM inference, these effects are amplified because activation tensors dominate memory volume, and even small inefficiencies in tiling or reuse propagate across multiple transformer blocks.
Therefore, maximizing end-to-end performance requires jointly reasoning about computation and data movement, rather than treating operator mapping and memory scheduling as separate problems.

Integrating a lightweight optimizer within its compiler stack can serve as an effective strategy.
Given model graph statistics and hardware constraints, the optimizer searches over tile sizes and stationary policies to find a configuration that minimizes estimated DRAM traffic while satisfying SRAM limits.
At runtime, the deterministic scheduling forces the selected dataflow, ensuring predictable latency and high utilization even under sub-megabyte on-chip memory.

In this paper, we propose \textit{TriGen}; a new \ac{NPU} architecture and optimizations such as corresponding neural network design and compilation methods for accelerating LLMs in environments on on-device settings.
TriGen is a software and hardware co-design architecture aimed at efficient resource utilization, designed to achieve optimal acceleration for \acp{LLM}, even in scenarios with limited resources. 
Notably, to the best of our knowledge, TriGen is the first attempt to apply \ac{MX} number system in an on-device environment, and it distinguishes itself from prior research by enabling the end-to-end execution of the entire model, addressing a critical gap in the field.

\vspace*{0.05in}
\noindent\textbf{Point 1: Adopting \ac{MX} and addressing its challenges:}
To enable 8-bit operations without sacrificing accuracy, TriGen introduces the \ac{MX} data type for the first time in an on-device context and proposes an architecture that seamlessly supports diverse mixed-precision configurations while addressing associated challenges.

\noindent\textbf{Point 2: Nonlinear operations acceleration using \ac{LUT}:} 
By leveraging a novel, fast, and accurate \ac{LUT} co-designed with software and hardware, TriGen enable the efficient processing of complex nonlinear operations such as softmax, \ac{SiLU}, and normalization without additional hardware such as vector processing unit and without any loss of accuracy.

\noindent\textbf{Point 3: Resource-aware dataflow mapping:} 
TriGen formulates dataflow mapping as a resource-aware optimization problem—co-optimizing tile sizes and stationary policies under compute resources, on-chip memory capacity and off-chip memory bandwidth constraints—to maximize arithmetic intensity and minimize data movement during on-device execution.

\vspace*{0.05in}
Collectively, the software-hardware co-design approach of TriGen, tailored for resource-constrained environments, achieves an average performance speedup of {2.73$\times$} and 52\% less memory transfer compared to the baseline \ac{NPU} design.

\section{Background and Related Work} \label{sec:background}
\begin{figure}[t]
    \centering
    \leftskip-0.04in
    \includegraphics[width=1.035\linewidth]{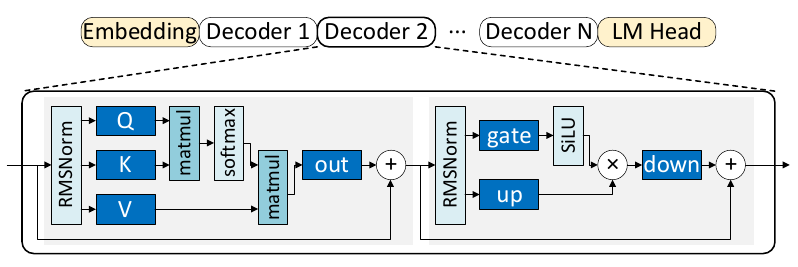}
    \caption{Structure of LLaMA3.2 LLM model
    }
    \label{fig:background_llm}
\end{figure}
\noindent{\textbf{Large Language Models.}}
In this paper, we choose Llama-3.2-3B \cite{touvron_arxiv2023} as a representative transformer based-LLM, which consists of multiple stacks of decoder layers.
Fig. \ref{fig:background_llm} illustrates a structure of Llama-3.2 decoder layer.
Input tokens (or sequence) are first converted into embeddings via embedding layer \cite{su_neurocomputing2024, vaswani_neurips2017} and the resulting embedding vectors are injected to the cascaded decoder layers.
A decoder layer consists of attention block and \ac{FFN}. 
Again, the attention block is composed of a series of \ac{RMSNorm}, $Q$/$K$/$V$ generation, attention score calculation, and output projection layers.
$Q$/$K$/$V$ generation layer performs \acp{matmul} between each pretrained parameter ($W_Q$, $W_K$, $W_V$) and input embeddings, and generate corresponding query ($Q$), key ($K$), and value ($V$).
During attention score calculation step, each $Q$, $K$, $V$ matrix is equally divided into $N_{head}$, number of heads, and compute similarity score using \ac{matmul} between $Q$ and $K$.
A softmax operation is applied to the resulting similarity score, and another \ac{matmul} is performed between softmax result and $V$.
Each output of multi heads are concatenated and inputted to the output projection layer, which performs projection using \ac{FC} layers.
The projected output goes through the \ac{FFN}, which consists of \ac{FC} layers and \ac{SiLU}.
Generated output token is used as the input token of the next decoder layer and the final output of the model is used as the input token of next iteration in a autoregressive manner.
\vspace*{0.04in}

\noindent\textbf{LLM Quantization on NPUs.}
Extensive research work have employed quantization to create lightweight DNN models with narrow bit width data formats.
Especially, quantization techniques tailored for \ac{LLM} have gained spotlight.
Most of those research focused on weight-only quantization, keeping the precision of activation unchanged and preserving accuracy \cite{dettmers_icml2023, frantar_iclr2023, lin_mlsys2024, lin_mlsys2025}.
However, using weight-only quantization lead to mismatching of weight and activation precision in terms of computation.
Frequent dequantization of weight into higher data precision (e.g., FP4 to FP16) incurs severe latency overhead.
To mitigate the dequantization overhead, weight-only \ac{LLM} accelerators introduced mixed-precision \ac{PE} design that can conduct \ac{MAC} operation between FP and INT data types \cite{fang_hpca2025, jang_hpca2024}.
Therefore, total cost (i.e., TOPS/${mm}^2$, or TOPS/W) of \ac{GEMM} operation is significantly reduced.

However, unlike the weight-only quantization, the existence of outliers whose values are extremely large compared to the others hinders the activation quantization.
To reduce the effect of outliers, weight-activation quantization research have introduced separately handling outliers with higher data precision while keeping the other activation in low precision \cite{dettmers_neurips2022, liu_iclr2025, xiao_icml2023}.
On the hardware side, outlier is computed with dedicated high precision \ac{MAC} units \cite{park_isca2018, lee_date2025} or custom number representations tailored for handling outliers \cite{guo_isca2023, li_tcas2024, ramachandran_isca2025}.

\section{Challenges and Motivations} \label{sec:motivation}
\noindent\textit{\textbf{Challenge 1: Low-Precision Activations in LLM Inference.}}
The use of low precision activations for LLM inference in resource-constrained environments, such as on-device NPU, provides an efficient way to improve the inference performance.
First of all, we can reduce the power consumption or the inference latency with low precision activations.
For example, with 8-bit activations and 4-bit parameters, a \ac{matmul} operator reduces the latency for the computation by almost half compared to the same operation with 16-bit activations and 4-bit parameters.
Also we can reduce the memory transfer for not only activations but also parameters.
There is no doubt about that the memory traffic of the activations would be reduced by their precision reduction, however, the reduction of the transfer size for the parameters looks not trivial but occurs frequently on an on-device \ac{NPU} whose on-chip memory is not sufficient to accommodate all of the related tensors.
 \begin{figure}[t]
     \centering
     \includegraphics[width=0.8\columnwidth]{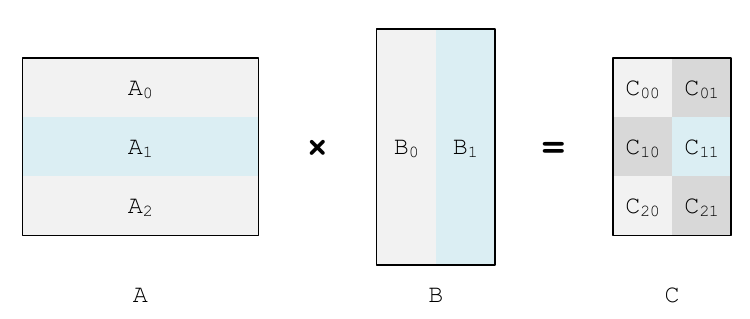}
     \caption{A tiling example of a \ac{matmul} $A \times B = C$}
     \label{fig:motivation_matmul}
     \vskip-0.1in
 \end{figure}
Fig. \ref{fig:motivation_matmul} illustrates an example of a \ac{matmul} with tiling. Sub-matrices $A_{i}$ and $B_{j}$ are read from the off-chip memory to the on-chip memory and used to make the result $C_{i,j}$ if the size of the on-chip memory is not large enough to store $A$, $B$, and $C$. Algorithm \ref{alg:matmul} explains the way in which most of efficient kernels iteratively run a sequence of sub-matrix multiplications. Please be noted that Algorithm \ref{alg:matmul} minimizes the data transfer for the \ac{matmul} by reading matrix $A$ once. Eq. \ref{eq:matmul_memory_transfer} shows the memory transfer size of the \ac{matmul} when $A$ is the stationary matrix.
\begin{equation}
transfer = \left\lceil \dfrac{rows(A)}{rows(A_{i})} \right\rceil \times size(B) + size(A)
\label{eq:matmul_memory_transfer}
\end{equation}
where $rows(A)$ is the number of the rows of $A$ and $size(A)$ is the size of $A$. The lower precision $A$ is, the larger sub matrix can be stored in the same size of on-chip memory, i.e., we can reduce the total transfer size of $B$ by decreasing the times of reading $B$. Fig. \ref{fig:eval_16_dram_traffic} demonstrates the comparative memory traffic of two inferences with 8-bit activation and 16-bit activation under identical optimization configuration. The inference with 16-bit activations consumes over 2 times more DRAM traffic comparing to the 8-bit inference.
\begin{algorithm}[t]
\caption{Sub-matrix Multiplication}
\label{alg:matmul}
\begin{algorithmic}[1]
\REQUIRE Inputs matrices : $A$, $B$
\ENSURE Matrix multiplication with input matrices
\FORALL{\textbf{each} $A_{i}$ \textbf{in} $A$}
\STATE read $A_{i}$ from off-chip memory
\FORALL{\textbf{each} $B_{j}$ \textbf{in} $B$}
\STATE read $B_{j}$ from off-chip memory
\STATE $C_{i,j} = A_{i} \times B_{j}$
\STATE write $C_{i,j}$ to off-chip memory
\ENDFOR
\ENDFOR
\STATE \textbf{return} $C$
\end{algorithmic}
\end{algorithm}
%
\begin{table}[h]
    \centering
    \caption{PPL Comparison Across Different Data Types}
    \label{tab:ppl_diff_data_type}
    \leftskip-0.024in
    \begin{tabular}{>{\centering\arraybackslash} m{0.595in}|
                    >{\centering\arraybackslash} m{0.48in}|
                    >{\centering\arraybackslash} m{0.52in}|
                    >{\centering\arraybackslash} m{0.53in}|
                    >{\centering\arraybackslash} m{0.53in}}                                                                                                               \hline
                   & \multicolumn{4}{c}{\textbf{Precision (Activation/Weight)}} \\ \hline
    \multirow{2}{*}{\textbf{Model}} & FP16   & INT16  & UINT8     & UINT8       \\ 
                                    & /UINT4 & /UINT4 & /UINT8    & /UINT4      \\ \hline
    Llama3.2-3B                     & 9.36   & 9.51   & 124912.73 & 125284.01   \\
    OPT-1.3B                        & 15.23  & 15.31  & 15905.23  & 21753.06    \\ \hline
    \end{tabular}
\end{table}

While it offers potential benefits in efficiency, the inherent accuracy degradation caused by de-quantization remains a critical issue, often compromising overall inference performance.
Table \ref{tab:ppl_diff_data_type} illustrates the comparative impact of varying precision levels on inference accuracy, as measured by \acp{PPL}.
A significant and abrupt degradation in \acp{PPL} is observed when precision is reduced from 16-bit to 8-bit through quantization.
This explains why 16-bit precisions for activation remains prevalent in industry applications~\cite{Chen_sosp2025, qualcomm_2025, apple_2024}, despite extensive academic research into low-precision alternatives.
Moreover, the adoption of low-precision data continues to present significant challenges in accuracy-critical tasks, potentially resulting in outcomes that are considered unacceptable.

In on-device NPU architectures, where SRAM capacity is inherently limited \cite{ethos-u85, hailo_2025}, underutilization of computational capacity (\ac{TOPS}) persists as a major challenge \cite{lennart_arxiv2025}.
Concurrently, inefficient resource allocation strategies can lead to increased DRAM access frequency, exacerbating bandwidth contention and energy inefficiency. In this context, the adoption of low-precision data, while preserving high accuracy, emerges as a critical enabler, offering the flexibility needed to address these computational and memory constraints.
\vspace{0.03in}
\begin{mdframed}
\textbf{Motivation 1.} Achieving high-accuracy inference with 8-bit low-precision quantization remains unresolved yet pivotal research challenges for NPU architectures subject to stringent resource constraints.
\end{mdframed}
\vspace{0.05in}

\noindent\textit{\textbf{Challenge 2: Acceleration of Nonlinear Operations.}}
\begin{figure}[t]
    \centering
    \leftskip-0.06in
    \includegraphics[width=1.03\linewidth]{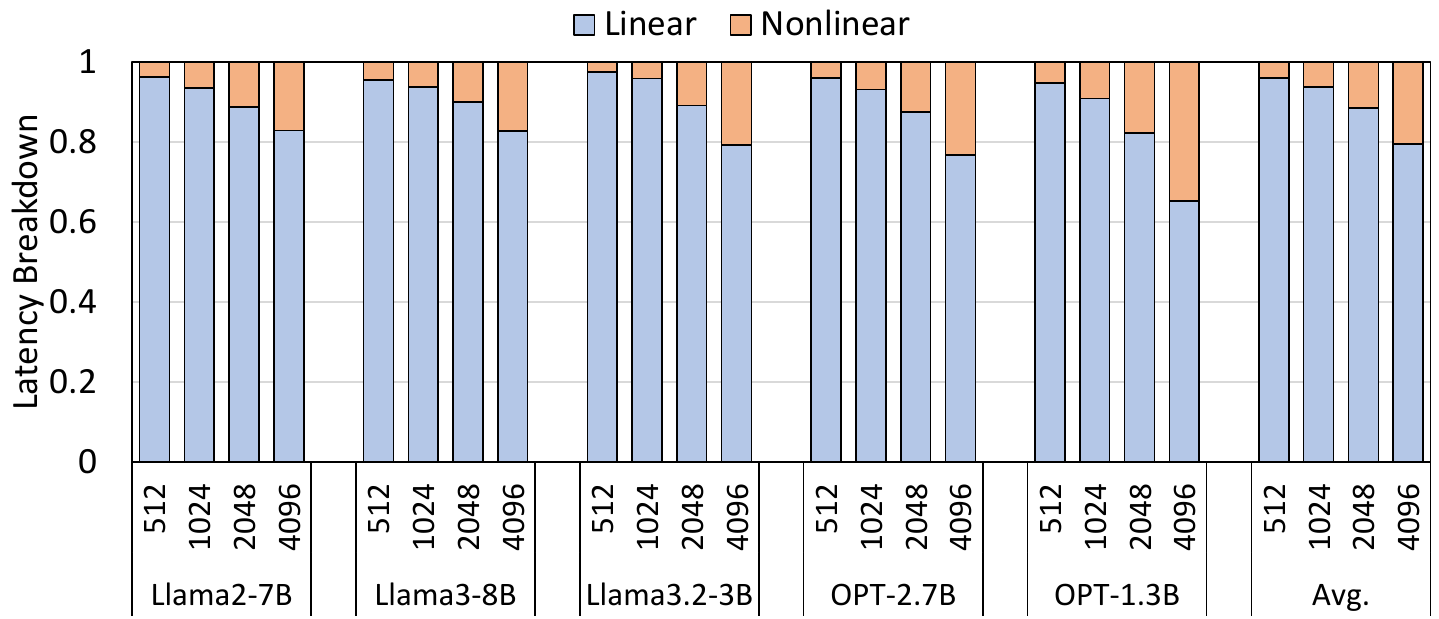}
    \caption{Latency breakdown of LLMs according to the input sequence length from 512 to 4096. 
             As the input sequence becomes longer, the portion of nonlinear function increases quadratically.
             When the input sequence length is 4k, nonlinear operations account for 20.5\% and of end-to-end execution time on average.
    }
    \label{fig:motivation_breakdown}
    \vskip-0.1in
\end{figure}

Fig. \ref{fig:motivation_breakdown} demonstrates the latency breakdown of end-to-end execution of five \acp{LLM}, Llama family \cite{touvron_arxiv2023} and OPT family \cite{zhang_arxiv2022} that are available on on-device \ac{NPU}.
Latency is measured varying the input sequence length from 512 to 4096.
As shown in Fig. \ref{fig:motivation_breakdown}, the portion of nonlinear operation increases quadratically according to input sequence length because of the attention module that is one of the core architectures of LLMs.
On average, the portion of nonlinear operation increases from 4.9\% to 20.5\% according to input sequence length.

Considering that input sequence length is growing (i.e., 2K, 4K input sequence), nonlinear operations will become major bottleneck for \ac{LLM} acceleration.
Simply focusing on speeding up only linear layers does not provide an effective solution for acceleration of \acp{LLM}.
\vspace{0.03in}
\begin{mdframed}
\textbf{Motivation 2.} 
As input sequences becomes longer, the portion of nonlinear operation grows quadratically.
Simply increasing computational throughput of linear operation (i.e., \ac{GEMM}) does not help for end-to-end performance speedup of \ac{LLM}. 
\end{mdframed}
\vspace{0.05in}

\noindent\textit{\textbf{Challenge 3: Memory-Constrained Operator Mapping.}}

Traditional operator mapping techniques for NPU, such as tiling, stationary data placement, and operation temporal ordering, are often optimized independently. 
However, in resource-limited environments such as on-device, this fragmented approach fails to address the interdependence of these factors. 
These elements exhibit a complex interplay; neglecting any one component risks resource contention, bottlenecks, and performance degradation.
Fig. \ref{fig:6_eval_dataflow_tile} illustrates the performance deviation resulting from improper dataflow mapping.

\vspace{0.03in}
\begin{mdframed}
\textbf{Motivation 3.} 
Efficient resource allocation must address the interplay of dataflow techniques to balance computational efficiency, memory bandwidth, and power consumption.
\end{mdframed}

\section{Architecture} \label{sec:architecture}
We propose \ac{NPU} architecture named TriGen that enables: i) low-precision \ac{LLM} inference with native \ac{MX}INT8 format support, ii) seamless acceleration of nonlinear function without \ac{SFU} or companion vector processor such as \ac{DSP}, and iii) efficient operator mapping that minimizes memory traffic while maximizing \ac{DLA} core utilization.

\begin{figure}[t]
    \centering
    \includegraphics[width=1.0\linewidth]{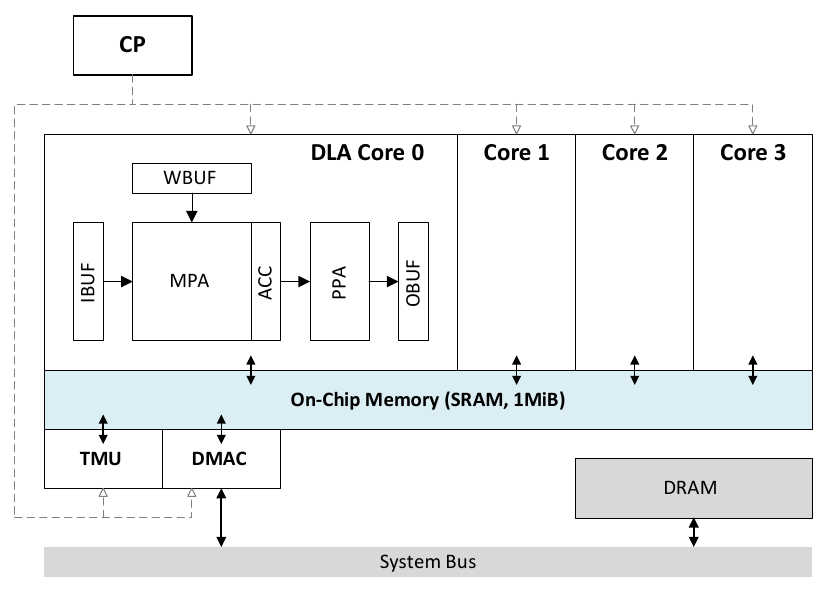}
    \caption{Overview of TriGen architecture}
    \label{fig:architecture_overview}
\end{figure}
\subsection{Architecture Overview} \label{subsubsec:trigen_overview}
Fig. \ref{fig:architecture_overview} illustrates overall architecture of TriGen.
An \ac{NPU} contains \ac{CP}, four \ac{DLA} cores, \ac{TMU}, and 1MiB global buffer (on-chip memory, SRAM).
\Ac{CP} is a lightweight RISC-V processor that controls and manages other components.
\Ac{CP} performs fetch, decode, and issue of instructions and offloads them to the related function unit.

A \ac{DLA} core has three buffers for inputs and output operand; IBUF, WBUF, and OBUF.
Dot product operations are conducted by \ac{MPA}. \ac{MPA} has 32 arrays and each array consists of 32 multipliers and one 32:1 adder tree.
Each array performs \ac{MAC} operations between two input vectors with length of 32 and therefore one \ac{MPA} performs 32$\times$32 \ac{MAC} operations at once.
The output result of \ac{MPA} is accumulated in \ac{ACC}.
\Ac{PPA} processes activation function such as ReLU, SiLU by referring to \ac{LUT}.
Also, if necessary, \ac{PPA} can perform quantization of output activation using rescale logic.

\Ac{TMU} is responsible for handling tensor manipulation, including, transpose, split, and concatenate of tensors.

\subsection{Supported Data Types in TriGen} \label{subsubsec:trigen_datatype}
TriGen facilitates the handling of input and output data types such as UINT4, UINT8, INT16, along with MXINT8, and their mixed precisions.
By leveraging integer data types for each operand, TriGen performs \ac{MAC} operation as integer addition and multiplication, which are significantly efficient in both latency and energy consumption \cite{horowitz_isscc2014, luo_ipdps2024}.
Furthermore, TriGen incorporates \ac{FI32} as an intermediate data type within the architecture's data flow.
By extending integer portion of MXINT8 data, FI32 format is capable of maintaining 24 bits of precision in the fractional part with an implicit scale factor.

\begin{figure}[h]
    \centering
    \includegraphics[width=1.0\linewidth]{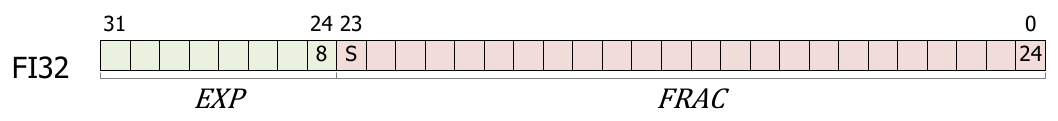}
    \caption{FI32 data type of TriGen
    }
    \label{fig:architecture_fi32}
\end{figure}
Fig. \ref{fig:architecture_fi32} illustrate structure of FI32 data format.
FI32 data format comprises two components: an 8-bit exponent (\texttt{EXP}) and a 24-bit integer fraction (\texttt{FRAC}) parts.
A real number can be expressed in FI32 format as follows:
\begin{equation}
    real\_number = FRAC \times implcit\_scale \times 2^{EXP} 
\label{eq:eq_fi32}
\end{equation} 
where $implicit\_scale$ is $2^{-22}$. 
FI32 and its variants offers a simple and intuitive representation compared to floating-point formats who needs a type conversion with sign for the accumulated results of TriGen.
Despite its simplicity, FI32 maintains high precision in the significant digits, making it an efficient data representation method.
FI32 is primarily utilized as an intermediate data type within the data flow between architectural modules. 
Additionally, FI32 is employed for external inputs and outputs from/to memory, such as bias (\texttt{BIAS}), partial sum (\texttt{PSUM}), and channelwise rescale (\texttt{CWQ}), where maintaining higher precision is essential.

\subsection{MAC Processing Array (MPA)} \label{subsubsec:trigen_mpa}

The \ac{MPA} comprises a $32\times32$ \ac{MAC} array architecture. Designed to support mixed-precision computation, the MPA incorporates an exponent (EXP) component in its data structure, enabling simultaneous processing of both \ac{MX} data and integer type data.
When processing integer data, the exponent component is set to 127 due to minus bias, while for MX-type data, the exponent and integer components are processed in parallel.
The EXP components are aggregated through summation operations, while the INT components undergo multiply-accumulate operations implemented via an adder tree structure.

The ACC unit performs the aggregation of partial sums where it firstly identifies the optimal common exponent value between the partial sum from the previous stage and the newly generated values from the MAC array, aligns the INT components based on this common EXP, and then aggregates two INTs to generate the intermediate data FI32 with EXP, which serves as another partial-sum value. 
TriGen architecture employs 64 \ac{ACC} registers for partial sum storage in one array (32 \acp{MAC}). To enable parallelized and efficient execution, it processes each operation (\ac{ISA}) by decomposing computations into multiple \textit{commands}, each handling 64 input row vectors.

The \texttt{TMATMUL}, which is a representative operation conducted in \ac{MPA}, has two input matrices IN0 and IN1, and it involves multiplying IN0 and transposed IN1. Let's consider the shapes of IN0 and IN1 as $(256 \times 64)$ and $(128 \times 64)$, respectively. The output's shape of \texttt{TMATMUL} would be $(256 \times 128)$. 
At first, a $(32 \times 32)$ sub matrix of IN1 are prefetched from on-chip memory and it remains stationary in \ac{MPA} during computation. After the prefetching, each $(1 \times 32)$ sub row vector of IN0 is read in every cycle and broadcasted to 32 arrays for performing dot product with the prefetched IN1 row vectors.
The MPA leverages a $(32 \times 32)$ MAC array to perform single-cycle dot product calculations, where one 32-element vector from IN0 and 32 same size vectors from IN1 are processed simultaneously.
The operation requires 4 commands due to the 64 partial-sum registers, accommodating IN0's 256-row dimension and each token undergoes partial-sum processing 4 times (aligning with the 128 out column dimension).
\begin{figure}[t]
    \centering
    \includegraphics[width=0.9\linewidth]{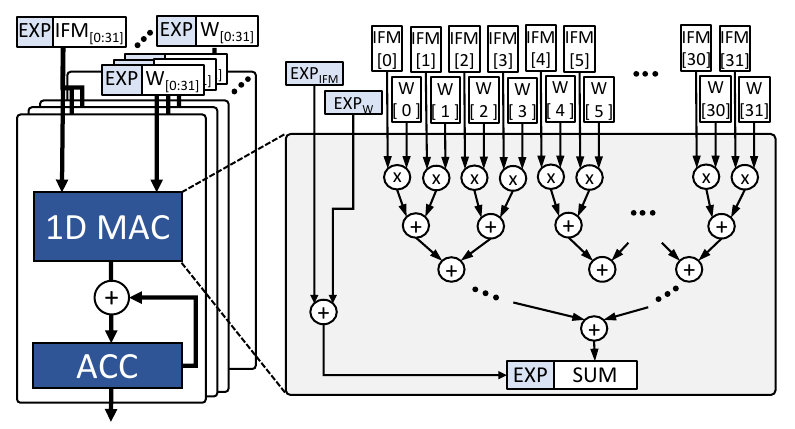}
    \caption{\ac{MPA} architecture of TriGen
    }
    \label{fig:architecture_mpa}
\end{figure}
\subsection{Post Processing Accelerator (PPA)} \label{subsubsec:trigen_ppa}
\begin{figure}[t]
    \centering
    \includegraphics[width=0.8\linewidth]{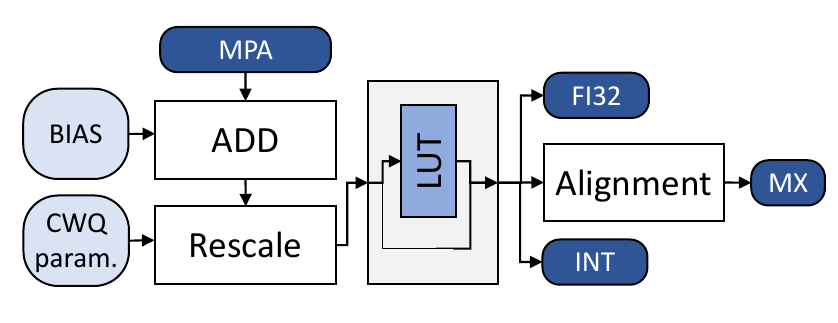}
    \caption{\ac{PPA} architecture in TriGen
    }
    \label{fig:architecture_ppa}
\end{figure}

The \ac{PPA} performs post processing using the intermediate results from the \ac{MPA} as the input. On the first stage, \ac{PPA} applies adding the bias and de-quantization on the input when it is necessary. For instance, if the parameter of a projection is INT4, RESCALE module de-quantizes the input suitable for the quantization configurations.
When necessary, the \ac{LUT} module is employed for processing nonlinear operations (e.g., exponential or SiLU), enabling efficient fusion with product-sum operations such as \texttt{TMATMUL} through this pipeline.
A detailed explanation of this process is provided in Section~\ref{subsec:lut}.

On the final stage, the \ac{PPA} performs output-specific operations tailored to each data type.
For FI32 outputs, it executes simple normalization to maximize significant digits.
For MX-type outputs, it identifies the maximum exponent value and performs corresponding bit-shifts on \texttt{FRAC} components to ensure proper alignment through its Alignment module.
For INT-type outputs, the PPA applies both zero-point adjustment and rounding operations to maintain data within the required bit-width constraints.

The \ac{PPA}, while primarily designed for the post processings, can also execute non-product-sum operations, such as element-wise \texttt{MUL}, \texttt{ADD}, and \texttt{RESCALE}, through control signal reconfiguration.

\subsection{Lookup Table (LUT)} \label{subsec:lut}
\begin{figure}[t]
    \centering
    \includegraphics[width=1.0\linewidth]{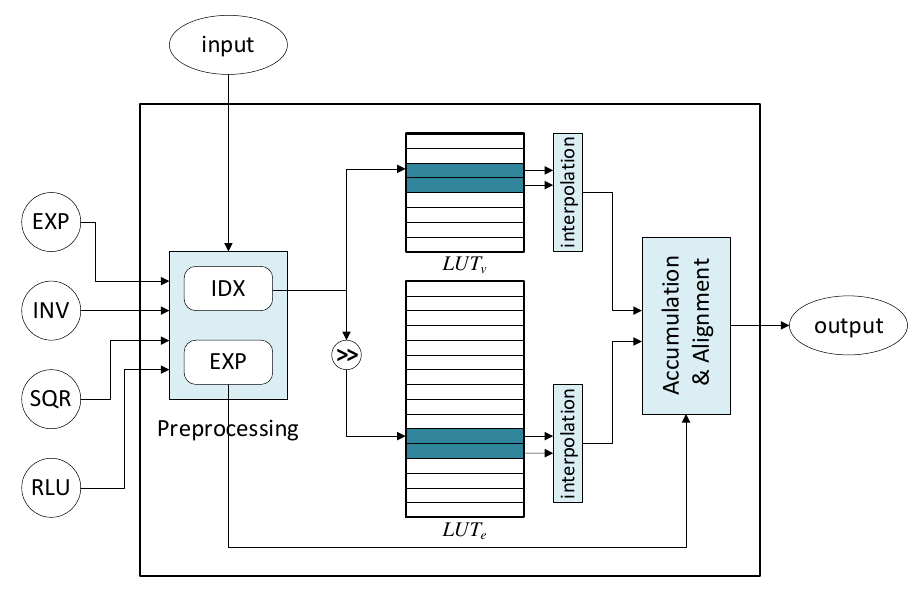}
    \caption{Overview of LUT architecture}
    \label{fig:lut_overview}
    \vskip-0.1in
\end{figure}
Fig. \ref{fig:lut_overview} illustrates overall architecture of \ac{LUT}. The preprocessing block computes the indices for retrieving values from the lookup tables and determines the shift amount for the table outputs. The examples below shows how pre-processing block works for \texttt{inverse square root} and \texttt{SiLU} operators.

\begin{itemize}
\item \textbf{\texttt{inverse square root}:} 
  $f(x) = \frac{1}{\sqrt{x}}$ is computed by decomposing $x$ as $x = 2^{e_x} \cdot m_x$, where $e_x$ is the exponent of $x$ and $1 \leq m_x < 2$. The approximation is given by:
\begin{equation}
    f(x) = 
    \begin{cases} 
    2^{-e_x/2} \cdot \frac{1}{\sqrt{m_x}} & \text{if } e_x \text{ is even}, \\
    2^{-(e_x+1)/2} \cdot \frac{1}{\sqrt{m_x/2}} & \text{if } e_x \text{ is odd}.
    \end{cases}
\label{eq:lut_isqr}
\end{equation}
  The lookup table for the inverse square root function contains values within the range $[\frac{1}{\sqrt{2}},\sqrt{2})$.
\item \textbf{\texttt{SiLU}:} The SiLU function $f(x) = \frac{x}{1 + e^{-x}}$ is computed by decomposing $x$ as $x=2^{e_x} \cdot m_x$, where $e_x$ is the exponent and $1 \leq m_x < 2$. Due to the asymptotic convergence of SiLU to ReLU for large \( |x| \), where $K$ is a predefined constant integer based on accuracy requirements, the approximation is given by:
\begin{equation}
    f(x) = 
    \begin{cases} 
    \frac{x}{1 + e^{-x}} & \text{if } e_x \leq K, \\
    2^{e_x - K} \cdot f(2^K \cdot m_x) & \text{if } e_x > K.
    \end{cases}
\label{eq:lut_silu}
\end{equation}
  The lookup table retains values of $f(2^K \cdot m_x)$.
\end{itemize}

\vspace{0.03in}
The value table and error table store precomputed values of nonlinear functions and their associated errors, respectively, enabling efficient computation of the function values in the operational model \cite{patent_lut}.
To achieve accurate but cost efficient solution, we utilize two tables: a compact table ($LUT_{v}$) with 16-bit entries and an expanded table ($LUT_{e}$) with 256 entries. The value table $LUT_{v}$ stores quantized values of the nonlinear function $f(x)$ at uniformly spaced input intervals, determined by the step size. The value interpolated $LUT_{v}(x)$ serves as an approximation of $f(x)$. The residual error $g(x) = f(x) - LUT_{v}(x)$ is stored in the $LUT_{e}$.

The interpolation block performs the linear interpolation on values retrieved from the related tables. The final result of the \ac{LUT} module is derived by summing the interpolated results of both ${LUT_v}$ and ${LUT_e}$, and adjusting the exponent of the resultant. 

\begin{table}[h]
\centering
\caption{Accuracy of nonlinear functions using \ac{LUT}}
    \begin{tabular}{>{\centering\arraybackslash} m{0.53in}|
                    >{\centering\arraybackslash} m{0.49in}|
                    >{\centering\arraybackslash} m{0.49in}|
                    >{\centering\arraybackslash} m{0.55in}|
                    >{\centering\arraybackslash} m{0.49in}}
    \hline
    \textbf{function} & \textbf{reciprocal} & \textbf{ISQR}\tnote{1} & \textbf{exponential} & \textbf{SiLU}\\ \hline
    \hline
    \textbf{input min} & \multicolumn{2}{c|}{1/1024} & \multicolumn{2}{c}{-8} \\ \hline
    \textbf{input max} & \multicolumn{2}{c|}{4096} & \multicolumn{2}{c}{64} \\ \hline
    \textbf{MAPE} & 8.397e-07 & 5.467e-06 & 2.023.e-05 & 1.626e-06 \\ \hline
    \textbf{MSE} & 2.434e-08 & 1.968e-07 & - & 7.344e-04 \\ \hline
\end{tabular}
\begin{tablenotes}
    \item[1] ISQR: invserse square root
\end{tablenotes}
\label{tab:lut_accuracy}
\end{table}

Table \ref{tab:lut_accuracy} shows the accuracy of our \ac{LUT} for various nonlinear functions. The input range for each function is discretized with a step size of 
$1/1024$ to ensure uniform sampling across the specified intervals. The input ranges are carefully selected to cover the operational domain of each function. The experiment demonstrates that our \ac{LUT} achieves exceptional accuracy in approximating nonlinear functions, with a \ac{MAPE} consistently below 0.1\% and a \ac{MSE} remaining under 
$1.0 \times 10^{-3}$. These results underscore the \ac{LUT}'s robustness and precision, making it a highly reliable solution for handling nonlinear-function approximations in practical applications.

\subsection{Support for Multi-NPUs} \label{subsubsec:trigen_multinpu}
Despite limited resources, modern on-device systems now aim to execute multiple models simultaneously to support diverse functionalities, leading to the growing adoption of multiple on-device NPUs for parallel processing. 
In this context, TriGen presents an efficient and scalable mechanism to distribute and executes the computation across multiple NPUs, enabling effective performance enhancement.

The core synchronization principle in TriGen's multi-NPU architecture ensures deadlock-free execution by blocking NPU progression until inter-NPU dependencies are resolved. 
To enable synchronized execution across multiple NPUs, the following architectural components are essential: 
(1) strategically embedded synchronization instructions with an appropriate Sync-ID within the instruction stream to demarcate coordination points; and (2) dedicated synchronization registers per NPU to autonomously manage and broadcast synchronization information.

When a specific NPU encounters a sync instruction during the execution of an instruction, it records the Sync-ID information, which represents the current stage being executed, in the Sync register and broadcasts it to all NPUs. 
At this point, from the perspective of the specific NPU, upon recognizing that the Sync-ID information from other NPUs has been broadcast (updated), it compares the value in its own Sync register with the Sync-ID information of the other NPUs to determine whether to proceed with execution. This lightweight approach efficiently handles synchronization in multi-NPUs where processing is divided into multiple parts.

\section{Software Optimizations} \label{sec:software_optimization}
Co-design of software and hardware is essential, especially in resource constraint environment such as on-device \ac{NPU}.
TriGen adopt software optimization techniques that i) remove unnecessary operation with operator fusion via mathematical equivalence and its derivation, ii) maximize \ac{PE} utilization while reducing DRAM access with software-defined dataflow and tiling strategy.

\subsection{Operator Optimization} \label{subsec:optimization_operator_fusing}

\noindent\textbf{Fusing transpose operation.}
Transposition of a matrix is a memory-bounded operation, and therefore can be a major bottleneck in accelerating matrix multiplication.
Moreover, when engaging \ac{MX} data format to activations, we cannot transpose the activation tensor as is since naive transposition of a matrix does not align with the axis of shared exponents and this causes severe accuracy drop.
Thereby, transposition of \ac{MX} type activation requires additional computations, 1) the previous layer makes result with a higher precision activation instead of \ac{MX} type activation, 2) the higher precision activation is transposed without loss of accuracy, and 3) the transposed higher precision activation is converted to \ac{MX} type activation.

Especially in \ac{LLM}, transpose operation is required in attention, where attention score is obtained as $softmax(QK^T)\times V$.
Since TriGen performs matrix multiplication in a transposed manner using \texttt{TMATMUL}, $Q\times K^T$ can be performed by single \texttt{TMATMUL} without additional transpose operation, i.e., $Q\times K^T = Q\times^T K$.
On the other hand, for the multiplication of attention score $A$ ($= softmax(QK^T)$) and $V$, $V$ should be transposed, i.e., $A\times V = A\times^T V^T$ such that output result of \texttt{TMATMUL} is equivalent to that without transposition.
To mitigate the overhead from explicit transposition of \ac{MX} format tensor, we introduce seamless fusion of $V^T$ into later calculation by formulation of mathematical equivalences. Eq. \ref{eq:v_projection} describes projection $X$ onto $V$.
\begin{equation} \label{eq:v_projection}
    V = X \times W_V
\end{equation}
$W_{V}$ is quantized as INT4 in usual LLM inferences on commercial products and, with applying the quantization, Eq. \ref{eq:v_projection} can be rewritten as below:
\begin{equation} \label{eq:v_quantization_2}
    V = X \times W_{INT} \times S_W
\end{equation}
where $S_W$ is per channel scale factor which is a tensor not a scalar value.
Applying transposition on both side, Eq. \ref{eq:v_quantization_2} becomes:
\begin{equation} \label{eq:v_quantization_3}
    V^T = {(X \times W_{INT} \times S_W)}^T = {S_W}^T \times {W_{INT}}^T \times X^T
\end{equation}
Please be noted the positions difference between $S_{W}$ in Eq. \ref{eq:v_quantization_2} and ${S_W}^T$ in Eq. \ref{eq:v_quantization_3}. General \acp{NPU} has rescaling block in its post processing module (\ac{PPA} for TriGen), that is, $\times S_{W}$ is fused into the previous \ac{matmul} $X \times W_{INT}$, however, ${S_W}^T \times$ cannot be fused and requires an additional \ac{matmul} operation.
We can express the result $Y$ of attention layer as:
\begin{equation} \label{eq:y_1}
    Y = A \times V = A \times (V^T)^T
\end{equation}
where $A$ is result of $softmax(QK^T)$. By putting $V^T$ of Eq. \ref{eq:v_quantization_3} into Eq. \ref{eq:y_1}, we obtain $Y$  as:
\begin{equation} \label{eq:y_2}
{\small
\begin{aligned}
    Y   &= A \times ({S_{W}}^T \times {({W_{INT}}^T \times X^T))}^T \\
        &= A \times {({W_{INT}}^T \times  X^T)}^T \times S_{W}
\end{aligned}
}
\end{equation}
Consequently, we can reformulate the fused transposition of $V$ as below:
\begin{equation} \label{eq:y_3}
    Y = (A \times^T {({W_{INT}}^T \times^T X)}) \cdot S_W
\end{equation}
As described in Eq. \ref{eq:y_3}, attention can be calculated as two consecutive \texttt{TMATMUL} operations.
Transposition of projection parameter matrix $W_{INT}$ can be handled during compile time.
Consequently, seamless fusion of transpose operation enables TriGen to avoid unnecessary transpose operation during runtime, which otherwise severely cost redundant data movement to transpose a tensor.

\noindent\textbf{Coalesced masking operation.}
To eliminate the influence of non-relevant data, a mask operation is necessary and this is conventionally performed via an elementwise multiplication (MUL) operation, however, unfortunately, this brings inefficiency since elementwise operations cannot fully utilize \ac{MPA}.
To mitigate this inefficiency, TriGen employs a fused \acp{LUT} within \texttt{TMATMUL} operations with \ac{PSUM}, effectively bypassing the need for a dedicated \textit{MUL} operation for masking.
\acp{PSUM} serves to aggregate values into \texttt{TMATMUL}'s output, enabling direct manipulation of fused \acp{LUT} inputs.
This achieves targeted results (e.g., zero-masking) without requiring separate operations.

In the \ac{SDPA} process of a softmax layer requiring exponentiation, TriGen strategically manipulates \ac{PSUM} to achieve efficient computation.
For non-masked values, it sets \ac{PSUM} to zero, while for masked positions, it configures \ac{PSUM} to the minimum \ac{FI32} value.
This approach effectively produces a zero output from the \acp{LUT} during exponentiation, eliminating the need for additional masking operations while maintaining computational accuracy.
Additionally TriGen skips the sub-matrix multiplication when all of the output of the sub-matrix multiplication operation would be masked out.
Later, we will evaluate the impact of operator fusions in Section \ref{subsec:evaluation_end_to_end_perf}.

\noindent\textbf{QKV Projections in a Batch.}
TriGen employs a batched processing approach for $Q$/$K$/$V$ projections, enabling simultaneous computation of required projections, rather than sequential execution.
The head-splitting is performed before projection, enabling tailored attention computation with only the necessary data.
This head-splitting allows loading parameters (IN1) of partial $W_Q$, $W_K$ and $W_V$, allowing them to fit entirely in SRAM.
TriGen loads required $W_Q$, $W_K$ and $W_V$ simultaneously in an IN1-stationary manner, then processes activation tiles sequentially with those parameters, maximizing computational efficiency while minimizing memory access for both of activations and parameters, which could have occurred repeatedly.

\subsection{Dataflow and Tiling Strategy}
\label{subsec:optimization_tiling_and_scheduling}

In on-device environments with constrained on-chip memory capacity, efficiently managing and preparing the appropriate data in a timely manner is essential for achieving high performance.
To ensure deterministic yet efficient data delivery, \textit{stationary} and \textit{tiling} strategies generally are considered.
The determination of optimal stationary strategy and tile size involves maximizing MAC utilization and the count of data reuse while minimizing DRAM traffic.
Furthermore, the number of MAC units, such as DLA cores in TriGen, must be considered to achieve balanced workload distribution.

TriGen adopts a simple yet effective systematic approach to determine these parameters, as described in Algorithm~\ref{alg:tiling_strategy}.
\begin{algorithm}[h]
\caption{Stationary and Tiling Selection in TriGen}
\label{alg:tiling_strategy}
\begin{algorithmic}[1]
\REQUIRE Inputs matrices : $in0$, $in1$
\ENSURE Matrix multiplication with input matrices
\FORALL{\textbf{each} $stationary$ \textbf{in} $\{in0, in1\}$}
\STATE generate division candidates based on DLA cores
\STATE determine the minimum tile size per candidate
\STATE select candidates with the maximum MAC utilization and largest tile size
\STATE expands \textit{the size of tiles} till fitting SRAM capacity
\ENDFOR
\STATE \textbf{return} stationary and tile dimension with the minimum DRAM traffic.
\end{algorithmic}
\end{algorithm}

TriGen systematically explores two stationary methods (IN0: first operand and IN1: second operand) to maximize MAC utilization while minimizing DRAM traffic. 
TriGen begins by generating division candidates based on the number of DLA cores and determines the corresponding minimum tile size, ensuring sufficient computation to hide overheads of DRAM data movement.
TriGen prioritizes candidates with the highest MAC utilization and largest tile dimensions in current stationary since a larger tile of the stationary input reduces reading times of the other input as shown in Eq. \ref{eq:matmul_memory_transfer}.
For instance, under IN0 stationary, it first selects cases where IN0 has the largest possible tile dimension.
The tile sizes are then incrementally expanded along the row dimension until all tiles (IN0, IN1, and OUT) fit within SRAM capacity.
This process repeats for the IN1 stationary method.
Finally, TriGen selects the optimal one between the candidate of two stationary methods based on minimal DRAM traffic, achieving an efficient computation-memory trade-off for matrix multiplication.

\section{Case Study: Deploying Llama on TriGen NPU} \label{sec:case_study}
In this section, we demonstrate a use case example of deploying \ac{LLM} on proposed TriGen.
We chose Llama-3.2-3B \cite{touvron_arxiv2023} as a representative transformer-based \ac{LLM} for the case study.
Throughout the case study, we use MXINT8 input activation whose sequence length is 2048. Thus dimension of input activation is $2048 \times 3072$ ($row \times column$).

\subsection{Normalize Layer} \label{subsec:case_study_norm}
Llama model can be split into two major parts; attention module and \acf{FFN} and, at the front of them, \ac{RMSNorm} \cite{zhang_neurips2019} layers exist.
\begin{equation}
\bar{a_{i}} = \dfrac{a_{i}}{\mathrm{RMS(a)}} \times g_{i} \text{, where } {\mathrm{RMS(a)}}=\sqrt{\dfrac{1}{n}\sum_{i=1}^{n}{a_{i}}^2}
\label{eq:RMSNorm}
\end{equation}

To conduct \ac{RMSNorm} on huge input activation, TriGen slices the input activation into multiple tiles.
Tile size is configured to maximize \ac{PE} utilization by hiding DRAM traffic behind the computation.
Table \ref{tab:case_RMSNorm_SRAM_usage} shows on-chip memory usage breakdown when the rows number of the input tile is is set to 96.
\begin{table}[h]
    \centering
    \caption{RMSNorm On-chip Memory Usage}
    \label{tab:case_RMSNorm_SRAM_usage}
    \begin{tabular}{>{\centering\arraybackslash} m{0.45in}|
                    >{\centering\arraybackslash} m{1.9in}|
                    >{\centering\arraybackslash} m{0.5in}}
                                                                             \hline
    \textbf{Name} & \textbf{Description}            & \textbf{Size (KiB)} \\ \hline
                                                                             \hline
    IN0           & \raggedright input tile of current computation  & 297 \\ \hline
    IN1           & \raggedright input tile for next computation    & 297 \\ \hline
    OUT           & \raggedright out tile of current computation    & 297                 \\ \hline
    SQSUM         & \raggedright square sums of IN0 & $<$ 1               \\ \hline
    LUT           & \raggedright lookup table data for inverse square root     & $<$ 1               \\
                                                                             \hline
    Total         &                                 & 893                 \\ \hline
    \end{tabular}
\end{table}

TriGen \ac{ISA} includes the operations below to accelerate whole process of \ac{RMSNorm}.
\begin{itemize}
    \leftskip-0.15in    

    \item{\texttt{\uline{\textbf{MEAN\_SQUARE \{IN0, OUT\}}}}: \texttt{MEAN\_SQUARE} reads every row from IN0 and calculates square mean ($\sum {{a_{i}}^2}$) of each row. The type of OUT is FI32 for keeping the accuracy and the shape is reduced to ($2048 \times 1$).} 
    
    \item{\texttt{\uline{\textbf{LUT \{IN0, OUT, flags\}}}}: \texttt{\ac{LUT}} reads input as IN0 and refer to \ac{LUT} according to flag combination. Flag can include \texttt{INV} (reciprocal), \texttt{SQR} (square root), \texttt{EXP} (exponential) and \texttt{RLU} (for SiLU). For \ac{RMSNorm}, IN0 is the OUT of \texttt{MEAN\_SQUARE} and both of \texttt{INV} and \texttt{SQR} flags are on. The result is given as OUT and it is used in \texttt{RESCALE}.} 

    \item{\texttt{\uline{\textbf{RESCALE \{IN0, OUT, scale\}}}}: \ac{PPA} multiplies scale to IN0 tensor, where scale is FI32 and IN0 is MXINT8 type. IN0 is the input of \ac{RMSNorm} and the scale is the out tensor of the previous \texttt{LUT} instruction.}
\end{itemize}

\subsection{Attention Module} \label{subsec:case_study_attention}
The resulting input sequences from \ac{RMSNorm} is projected into each $Q$, $K$, and $V$.
The projections are matrix multiplication between input token tensor $X$ ($2048 \times 3072$) and the related parameter tensors $W_Q$ ($3072 \times 3072$), $W_K$ ($3072 \times 1024$), and $W_V^T$ ($1024 \times 3072$). As we explained in Section \ref{subsec:optimization_operator_fusing}, each head attention processes the projections with only the related parameters, thus the shape of a parameter tensor in a head attention is ($3072 \times 128$).
TriGen performs matrix multiplication using \texttt{TMATMUL} instruction.
\begin{itemize}
    \leftskip-0.15in

    \item{\texttt{\uline{\textbf{TMATMUL \{IN0, IN1, OUT, flags\}}}}: \ac{MPA} receives two inputs and performs transposed matrix multiplication between IN0 and IN1. Result is stored as OUT tensor.}

\end{itemize}

Resulting $Q$, $K$, and $V^T$ tensors go through attention mechanism. First, \ac{MPA} performs $Q \times^T K$ via \texttt{TMATMUL} instruction, then, softmax operation is applied to the out tensor.
Softmax requires the sum of exponential of input tensor (i.e., $\sum e^{x_i}$) and the inverse of the sum.
Similar to \ac{RMSNorm}, TriGen handles exponential via \texttt{LUT} instruction with \texttt{EXP} flag.
Then, summation is computed in \ac{PPA} via \texttt{MEAN} instruction.
\begin{itemize}
    \leftskip-0.15in
    \item{\texttt{\uline{\textbf{MEAN \{IN0, OUT, flags\}}}}: \ac{PPA} computes the summation of IN0 tensor elements along depth dimension and returns the resulting scalar value as out.}
\end{itemize}
$V^T$ is multiplied to the softmax result via \texttt{TMATMUL} instruction and $softmax(QK^T) \times^T V^T$ becomes final output of attention layer.

\subsection{Feed Forward Network} \label{subsec:case_study_ffn}
\Ac{FFN} consists of multiple \ac{MLP} layers; gate, up, and down projection, respectively.
Similar to attention layer, TriGen handles such projection operation via \texttt{TMATMUL} instruction.
For the gate projection, whose output goes into SiLU activation function, TriGen allows the \texttt{RLU} flag to be set on and SilU function is fused with the projection. At the final stage of \texttt{TMATMUL}, \ac{PPA} applies SiLU on the accumulated results from \ac{MPA} when the \texttt{RLU} flag is on. Please be noted that an accumulated result is intermediate data and it has high precision (FI32). Thanks to the high precision data and the advanced architecture of LUT, fused SiLU doesn't make accuracy loss. 
Output from up projection with fused SiLU goes through elementwise multiplication via \texttt{MUL} instruction.
\begin{itemize}
        \leftskip-0.15in
    \item{\texttt{\uline{\textbf{MUL \{IN0, IN1, OUT, flags\}}}}: \ac{MPA} receives two inputs and performs elementwise multiplication between two tensors IN0 and IN1. Result is returned as OUT tensor.}
\end{itemize}
Again, the result undergoes down projection \ac{MLP} via \texttt{TMATMUL}, and final residual operation is conducted using \texttt{ADD} instruction, which performs elementwise addition.
\begin{itemize}
        \leftskip-0.15in
    \item{\texttt{\uline{\textbf{ADD \{IN0, IN1, OUT, flags\}}}}: \ac{MPA} receives two inputs and performs elementwise addition between two tensors IN0 and IN1. Result is returned as out tensor.}
\end{itemize}

\section{Evaluation} \label{sec:evaluation}
\begin{table}[t]
    \centering
    \caption{TriGen NPU Configuration}
    \label{tab:evaluation_npu_config}
    \begin{tabular}{>{\centering\arraybackslash} m{1.1in}|
                    >{\centering\arraybackslash} m{1.55in}}
                                                             \hline
    \textbf{Configurations} & \textbf{Parameters}         \\ \hline
                                                             \hline
    Command processor       & single issue RISC-V @ 1 GHz \\ \hline
    \# of NPUs              & \textbf{1} / 2 / 4 / 8      \\ \hline
    NPU clock frequency     & 1 GHz                       \\ \hline
    MPA dimension           & 32 $\times$ 32              \\ \hline
    \# of DLAs per NPU       & 1 / 2 / \textbf{4}          \\ \hline
    \# of LUT entries       & $LUT_v$: 16, $LUT_e$: 256   \\ \hline
    DRAM bandwidth          & 32 GB/s                    \\ \hline
    Global buffer size      & 1 MiB       \\ \hline
    Data format             & MXINT8 (act.),
                              UINT4 (weight),
                              FI32 (intermediate value)   \\ \hline
    \end{tabular}
\end{table}
\noindent\textbf{Performance Modeling.} 
We model TriGen as a cycle-level simulator implemented in C++.
The latency obtained from simulator is verified against RTL result from Synopsys Design Compiler using 14nm technology node.
TriGen is configured to the parameters listed in Table \ref{tab:evaluation_npu_config}.
As a baseline configuration, we use one \ac{NPU} operating at 1 GHz.
The number of \acp{NPU} are varied from one to eight cores, and we present the trade-offs in implementing multi core \ac{NPU}.
Each \ac{NPU} is engaged with 32 $\times$ 32 \ac{MPA}.
DRAM bandwidth is set up to 32 GB/s and, in multi-\ac{NPU} setup, the average bandwidth for one NPU is \textsf{(32 / number of NPUs)}. 
Global buffer size is set to 1 MiB per \ac{NPU} and it can be varied.
We use MXINT8 data type for activation and UINT4 for weight.

\begin{table}[t]
    \centering
    \caption{LLMs Used in the Experiment}
    \label{tab:evaluation_workload_config}
    \begin{tabular}{>{\centering\arraybackslash} m{0.6in}|
                    >{\centering\arraybackslash} m{1.425in}|
                    >{\centering\arraybackslash} m{0.75in}}
                                                                                                                              \hline
    \textbf{LLMs} & \textbf{Quantization Scheme}                                      & \textbf{Dataset}                   \\ \hline
                                                                                                                              \hline
    Llama2-7B     & QuaRot \cite{ashkboos_neurips2024} + GPTQ \cite{frantar_iclr2023} &                                    \\
    Llama3-8B     & SpinQuant \cite{liu_iclr2025} + GPTQ \cite{frantar_iclr2023}      &  WikiText-2 \cite{merity_iclr2017} \\ 
    Llama3.2-3B   & SpinQuant \cite{liu_iclr2025} + GPTQ \cite{frantar_iclr2023}      &  C4 \cite{raffel_jmlr2020}         \\
    OPT-1.3B      & QuaRot \cite{ashkboos_neurips2024} + GPTQ \cite{frantar_iclr2023} &  PTB \cite{marcus_hlt1994}         \\
    OPT-2.7B      & AWQ \cite{lin_mlsys2024} + GPTQ \cite{frantar_iclr2023}           &                                    \\ \hline
    \end{tabular}
\end{table}
\noindent\textbf{Workload and Dataset.} 
Table \ref{tab:evaluation_workload_config} shows the list of \acp{LLM} and dataset used in the experiment.
We choose \acp{LLM} that are deployable within the resource limitations of on-device \acp{NPU}; Llama-2-7B, Llama-3-8B, Llama-3.2-3B \cite{touvron_arxiv2023}, OPT-1.3B, OPT-2.7B \cite{zhang_arxiv2022}.
We mainly use the language model tasks with WikiText-2 \cite{merity_iclr2017}, C4 \cite{raffel_jmlr2020}, PTB \cite{marcus_hlt1994} dataset.
Activation and weight of \acp{LLM} are quantized to each data formats using various quantization schemes and we choose the one that shows the best \ac{PPL} for each precision combination.
For instance, Llama-2-7B was quantized with QuaRot \cite{ashkboos_neurips2024} and GPTQ \cite{frantar_iclr2023} while Llama-3-3.2B used combination of SpinQuant \cite{liu_iclr2025} and GPTQ.

\subsection{End-to-End Latency Analysis} \label{subsec:evaluation_end_to_end_perf}

\begin{figure*}[t]
  \centering
  \includegraphics[width=0.9\textwidth]{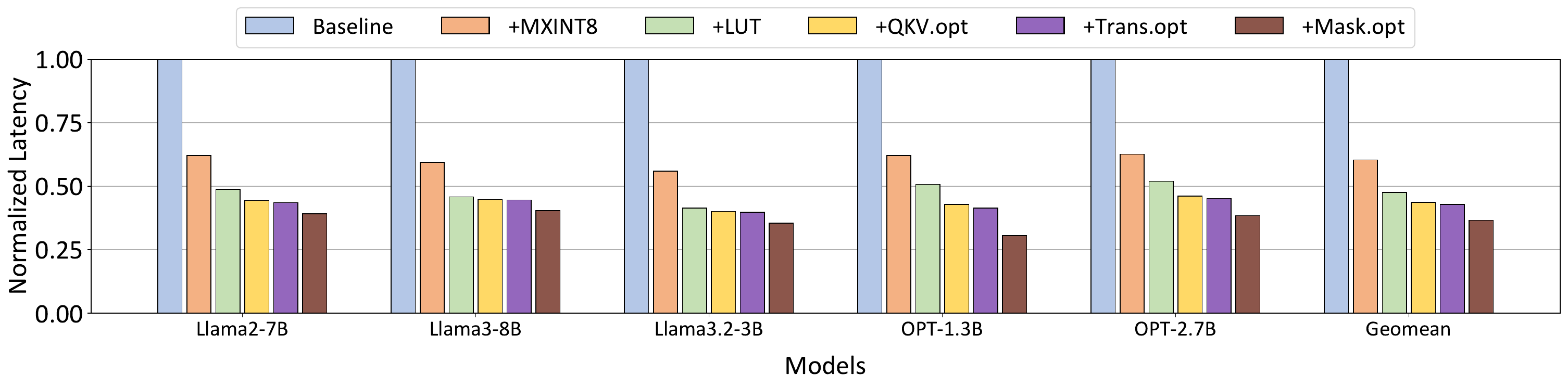}
  \caption{Normalized latency of TriGen during LLM inference.
           We measure the latency of proposed optimization scheme and demonstrate as a cumulative latency.
           On average, TriGen with all optimization scheme delivers 2.73$\times$ speedup over baseline.}
  \label{fig:prefill-latency}
  \vspace{-0.1in}
\end{figure*}

We evaluate the end-to-end latency of TriGen across five open LLMs (Table \ref{tab:evaluation_workload_config}) with sequence lengths 2048.
We compare the following design points of TriGen:
\begin{itemize}
    \leftskip-0.15in
    \item{\textit{Baseline:} Baseline models TriGen illustrated in Fig. 4 which performs \ac{LLM} inference using INT16/UINT4 bit quantization. Nonlinear operation is handled by special function unit, which incurs significant overhead.}
    \item{\textit{+MXINT8:} This configuration adopts MXINT8 data format to the input activation, and therefore reduces DRAM traffic and bandwidth.}
    \item{\textit{+LUT:} This configuration performs nonlinear operation using novel \acp{LUT} with negligible accuracy loss.}
    \item{\textit{+QKV.opt:} This configuration leverages batched $Q$/$K$/$V$ projection, which reduces DRAM accesses for input activation.}
    \item{\textit{+Trans.opt:} This configuration performs fusing transpose operation as discussed in Section V-A.}
    \item{\textit{+Mask.opt:} Built upon previous schemes, this configuration of TriGen replaces multiplication during masking operation with fused elementwise addition.}
\end{itemize}

\begin{figure}[t]
    \centering
    \includegraphics[width=0.85\columnwidth]{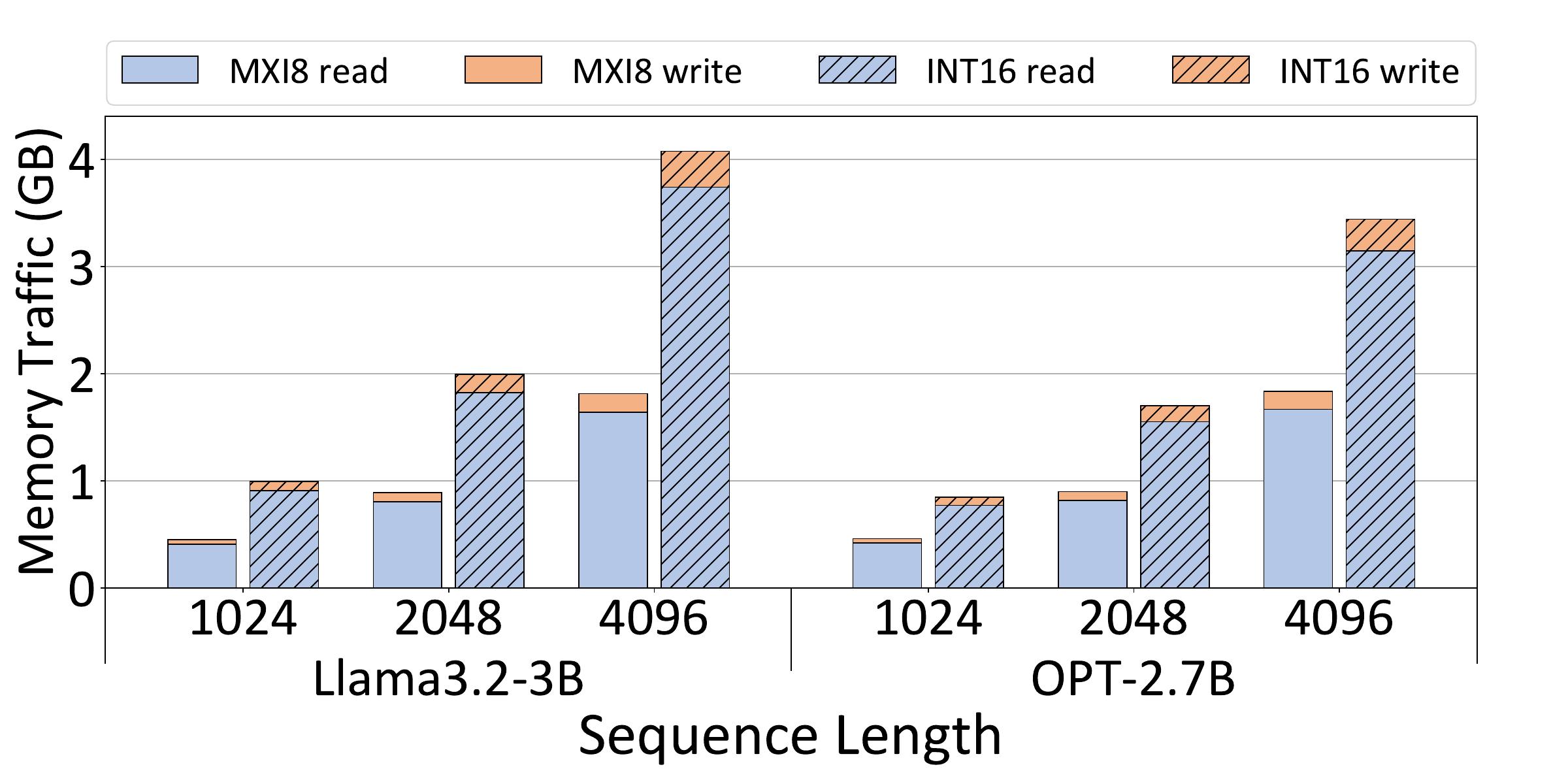}
    \caption{DRAM traffics with 16-bit activation. Numbers are normalized to that of 8-bit activation.}
    \label{fig:eval_16_dram_traffic}
    \vskip-0.1in
\end{figure}

Fig. \ref{fig:prefill-latency} demonstrates the effect of each TriGen optimization scheme on the overall latency of \ac{LLM} inference.
Overall, engaging MXINT8 format to activation offers 1.65$\times$ speedup.
This significant performance gain mainly comes from increased numerical throughput and reduced DRAM traffic.
Especially, the amount of DRAM traffic is diminished by approximately half as shown in Fig. \ref{fig:eval_16_dram_traffic}.

In the baseline configuration, nonlinear operation is computed by \ac{SFU}, which accounts for 15.2\% of total execution time.
Engaging \acp{LUT} for nonlinear operation (labeled as \textit{LUT}) further improves the performance by 26.8\% compared to \textit{MXINT8} scheme.
Furthermore, batching QKV projection (\textit{QKV.opt}) improves the performance by 9.1\%.
This technique is especially effective when hidden dimension is small.
For instance, OPT-1.3B has hidden dim of 2048 while that of Llama3.2-3B is 3072.
Small hidden dimension leads to expose of latency to load each activation and weight tile, since the amount of computation is not sufficient to hide the memory latency.
Consequently, OPT family benefits the most from the batching QKV.

\textit{Trans.opt} removes transpose operation of $V$ and replaces it with reverse ordered \ac{matmul}.
On average, \textit{Trans.opt} offers 1.77\% improvement over previous scheme.
Masking of \ac{SDPA} requires elementwise multiplication.
\textit{Mask.opt} technique substitutes the multiplication with addition as a bias after \ac{matmul}.
Therefore, this scheme can remove the elementwise multiplication and seamlessly performs masking operation.
Collectively, the propose techniques achieves 2.73$\times$ speedup over baseline TriGen design.

\subsection{Impact of Data Precision} \label{sec:precision_eval}
To examine how precision configurations influence model accuracy, we analyze multiple workloads using various combinations of activation and weight data types. 

\begin{table}[h]
    \centering
    \caption{PPL Measured on TriGen}
    \label{tab:eval_ppl}
    \leftskip-0.02in
    \begin{tabular}{>{\centering\arraybackslash} m{0.59in}|
                    >{\centering\arraybackslash} m{0.48in}|
                    >{\centering\arraybackslash} m{0.52in}|
                    >{\centering\arraybackslash} m{0.53in}|
                    >{\centering\arraybackslash} m{0.53in}}
                                                                                                    \hline
                                    & \multicolumn{4}{c}{\textbf{Precision (Activation/Weight)}} \\ \hline
    \multirow{2}{*}{\textbf{Model}} & MXINT8 & INT16  & UINT8     & UINT8     \\
                                    & /UINT4 & /UINT4 & /UINT8    & /UINT4    \\ \hline
    Llama2-7B                       & 6.07   & 6.24   & 40705.77  & 37447.06  \\
    Llama3-8B                       & 7.49   & 7.48   & 154756.14 & 168274.83 \\
    Llama3.2-3B                     & 9.37   & 9.51   & 124912.73 & 125284.01 \\
    OPT-1.3B                        & 15.98  & 15.31  & 15905.23  & 21753.06  \\ 
    OPT-2.7B                        & 12.97  & 13.03  & 12.77     & 13.44     \\ \hline
    \end{tabular}
\end{table}

Table \ref{tab:eval_ppl} illustrates the \ac{PPL} variations across different activation–parameter precision pairs.
The results reveal that model \textit{parameters} can be quantized down to 4 bits with negligible accuracy loss, however, reducing \textit{activations} to UINT8 causes substantial degradation in \ac{PPL}, even under the best choice of quantization algorithms described in \ref{tab:evaluation_workload_config}.
This indicates that activations require a precision level equivalent to roughly 16 bits of effective resolution to sustain the quality. 

Fig. \ref{fig:eval_16_dram_traffic} illustrates the comparison of DRAM traffics when employing either \acp{MX} or INT16.
While \ac{MX} is applied solely to activations, its adoption significantly impacts the number of actual parameter loading, resulting in around 50\% reduction in DRAM traffic compared to 16-bit implementations.

Consequently, TriGen adopts the \ac{MX} data representation, which provides activation precision comparable to INT16 while significantly reducing memory traffic and bandwidth consumption.
This design choice achieves a balanced trade-off between computational efficiency and accuracy preservation.

\subsection{Impact of Dataflow and Tiling} \label{subsec:evaluation_dataflow_tiling}

\begin{figure}[t]
    \centering
    \includegraphics[width=0.95\columnwidth]{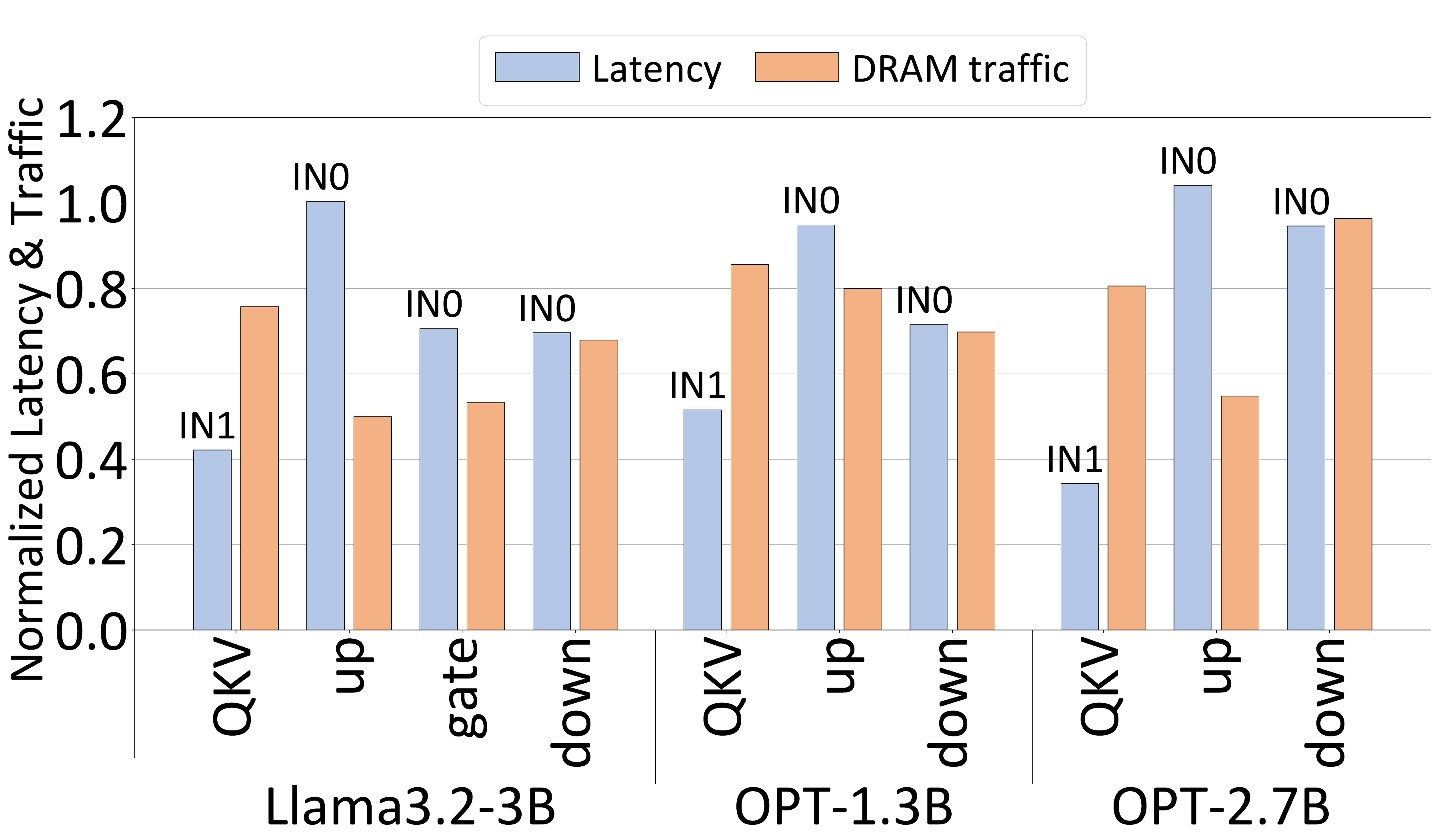}
    \caption{Normalized latency and DRAM traffics to selected stationary.}
    \label{fig:6_eval_dataflow_stationary}
\end{figure}

Based on Algorithm \ref{alg:tiling_strategy} (in Section \ref{subsec:optimization_tiling_and_scheduling}), TriGen determines optimal stationary configurations and tile dimensions.
Fig. \ref{fig:6_eval_dataflow_stationary} presents the comparative performance of the strategies.
For the Llama-3.2-3B model, TriGen empolys an IN1 stationary for $QKV$ projection and the latency and the memory traffic are reduced by 58\% and 24\% respectively comparing to the performance of IN0 strategy.
Through this targeted stationary selection, TriGen achieves superior performance. 
Although latency remains comparable for up and gate projections under both stationary in all model, memory bandwidth consumption is reduced by 20-50\%, directly contributing to enhanced overall performance particularly in terms of on power consumption.
\begin{figure}[t]
    \vskip-0.12in
    \centering
    \includegraphics[width=0.95\columnwidth]{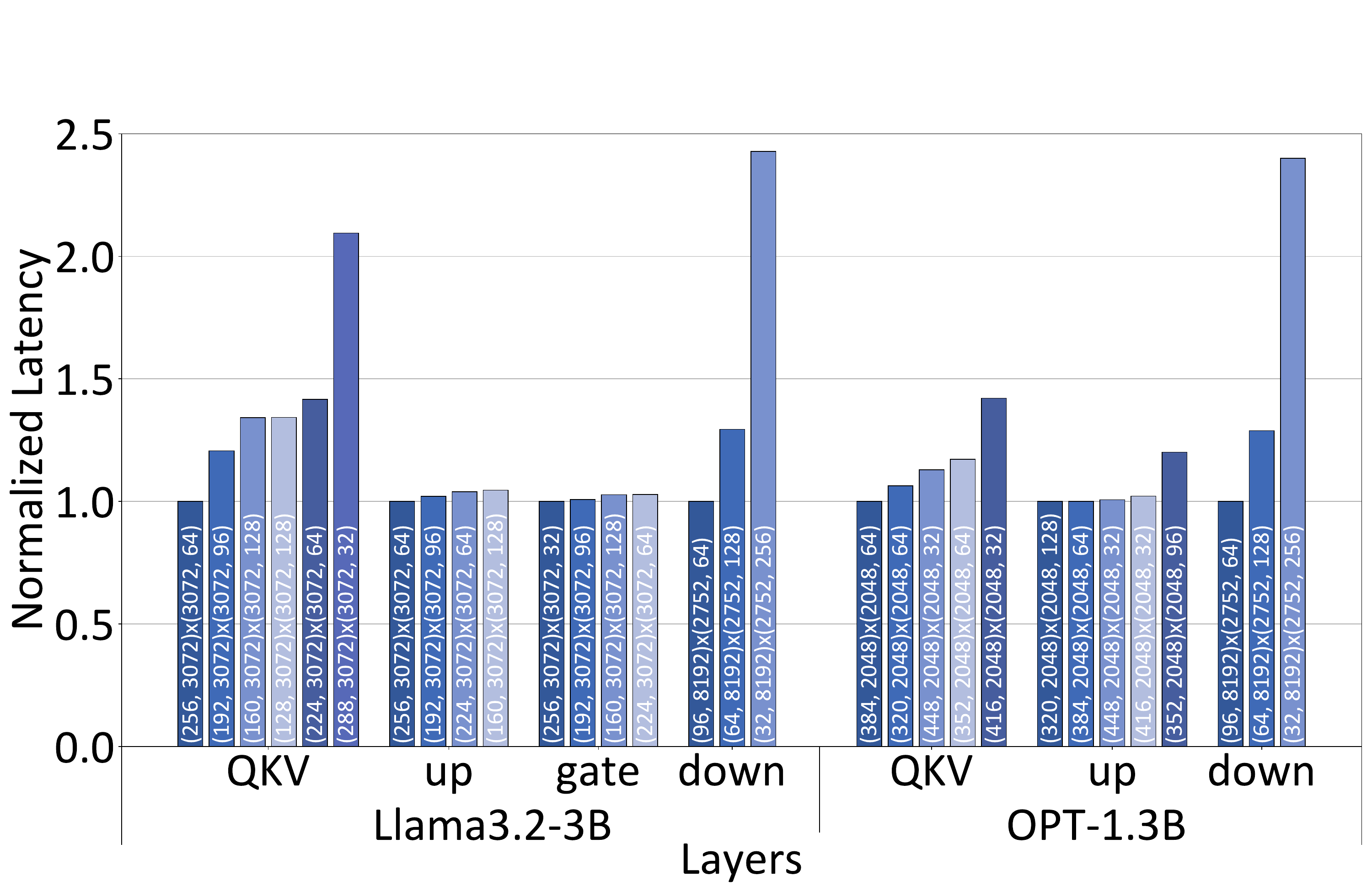}
    \caption{Normalized latency to selected tile dimension.}
    \label{fig:6_eval_dataflow_tile}
\end{figure}

Fig. \ref{fig:6_eval_dataflow_tile} compares normalized latency across candidate tiling configurations for diverse layers. Each bar represents latency normalized against the optimal tiling configuration for its respective layer. For instance, in the $Q$/$K$/$V$ projection of the Llama3.2-3B model, configuration ($(192, 3072) \times (3072, 96)$) shows 20.6\% latency overhead compared to the optimal ($(256, 3072) \times (3072, 64)$). On average, our proposed tiling strategy (Algorithm~\ref{alg:tiling_strategy}) achieves 12.5\% latency reduction over second-best configuration.

\subsection{Sensitivity and Scalability Analysis} \label{subsec:evaluation_scalability}
\noindent\textbf{Performance variation on multi-NPUs.}
TriGen achieves scalable multi-NPU performance by employing a low-overhead synchronization mechanism and optimized workload balancing through its dataflow and tiling strategy.
Each NPU comprises four DLA cores and 1MiB SRAM. Workload and system DRAM bandwidth are allocated evenly across all NPUs, ensuring balanced resource utilization and scalability.

\begin{figure}[t]
    \centering
    \begin{subfigure}[b]{0.6\columnwidth}
        \includegraphics[width=\linewidth]{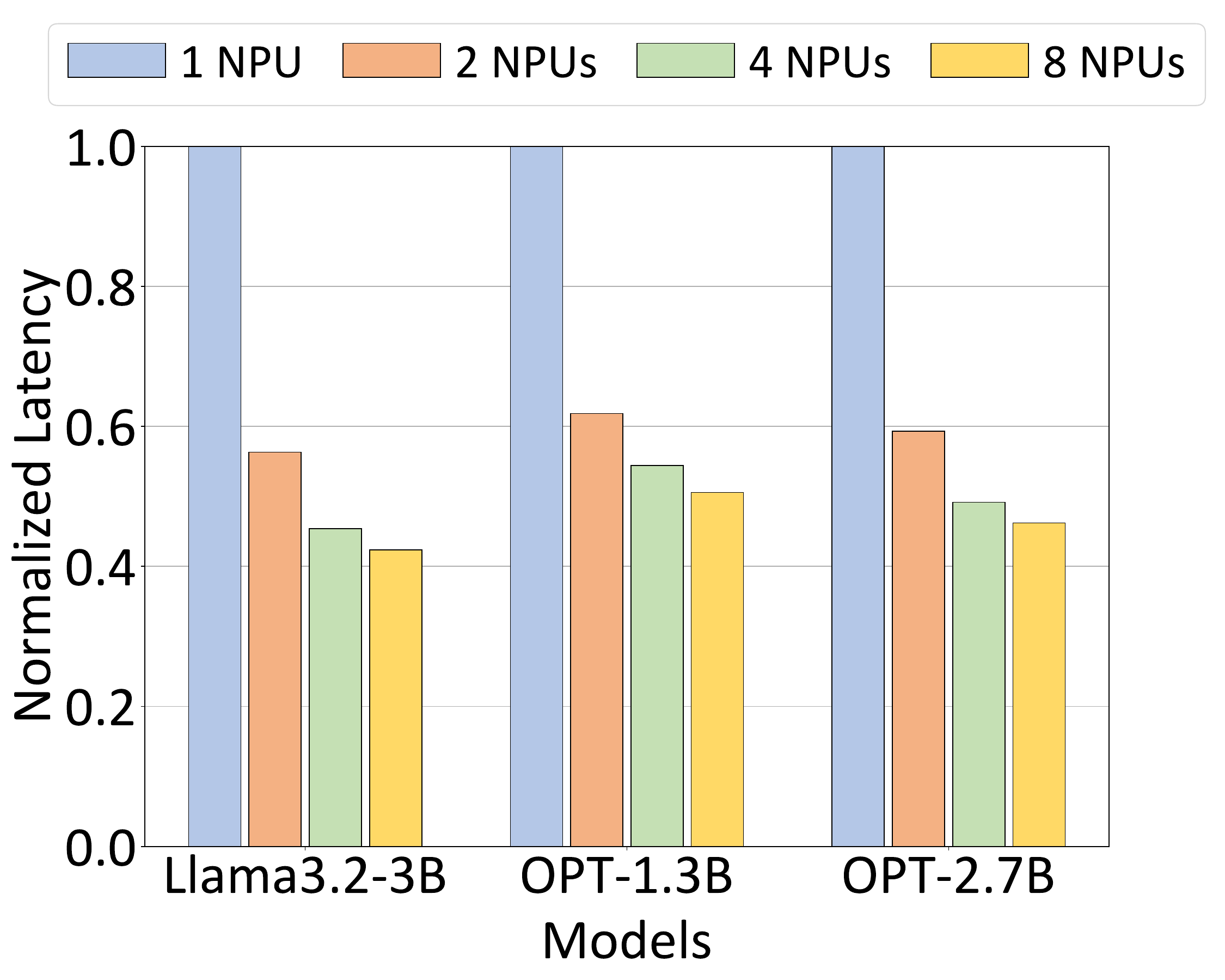}
        \caption{}
        \label{fig:6_eval_multi_npu_latency_1}
    \end{subfigure}
    \hspace{-0.7em} 
    \begin{subfigure}[b]{0.4\columnwidth}
        \includegraphics[width=\linewidth]{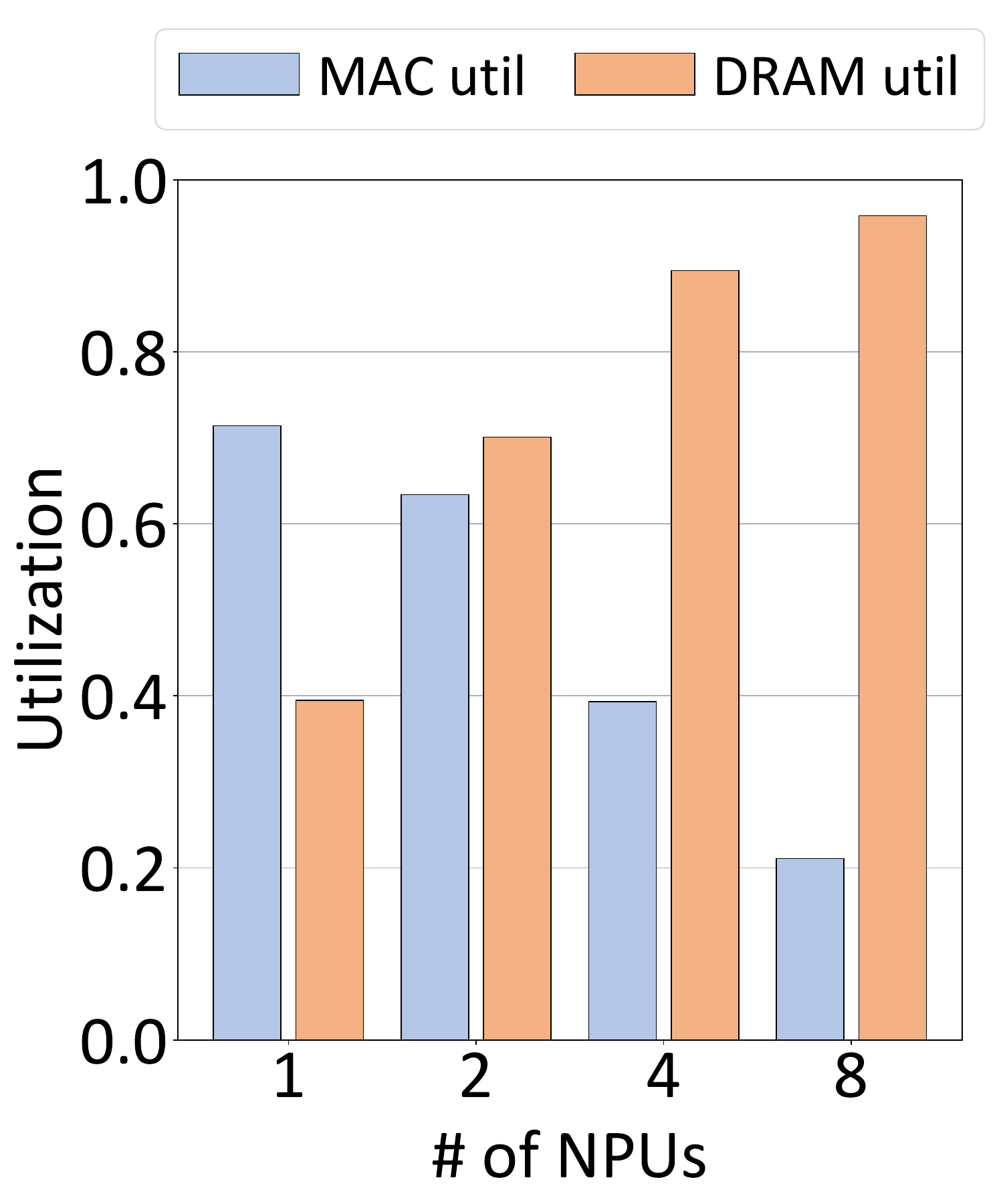}
        \caption{}
        \label{fig:6_eval_multi_npu_util_1}
    \end{subfigure}
    \caption{Normalized latency  varying the number of NPUs: (a) Latency comparison, (b) MAC and DRAM utilization.}
    \label{fig:6_eval_multi_npu_latency}
\end{figure}
\begin{figure}[t]
    \centering
    \includegraphics[width=0.95\columnwidth]{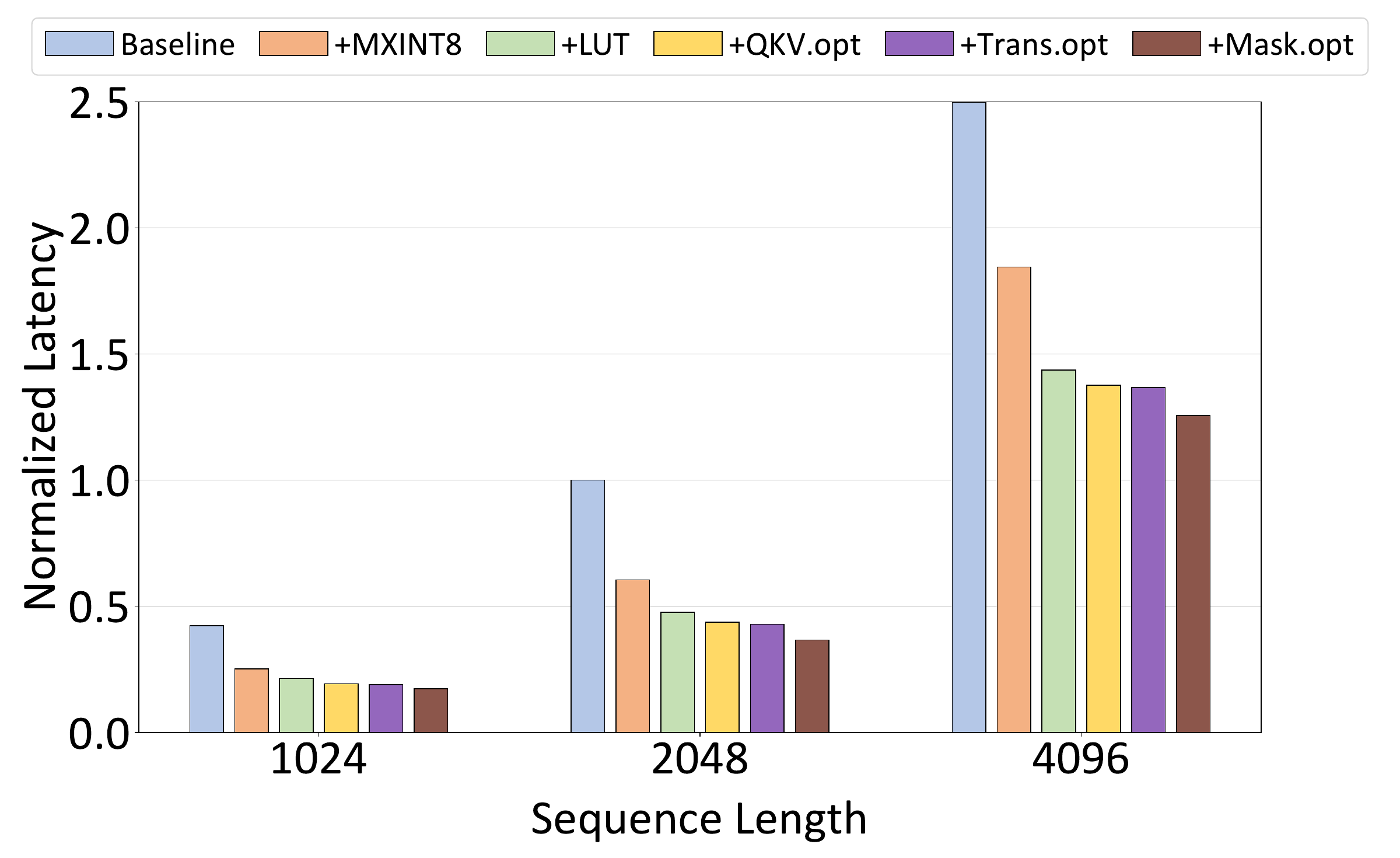}
    \caption{Normalized latency varying the length of sequence length. All bars are normalized to the baseline of 2K-tokens.}
    \label{fig:6_eval_input_sequence}
\end{figure}
Fig. \ref{fig:6_eval_multi_npu_latency} demonstrates TriGen's consistent scalability up to 4 \acp{NPU} under a 4K-token sequence length using the Llama-3.2-3B model. 
At 8 NPUs, however, scalability exhibits a moderate reduction due to memory bandwidth constraints. As the number of \acp{NPU} increases, the allocated DRAM bandwidth to an \acp{NPU} decreases since the total DRAM bandwidth of the system is fixed. Therefore, the characteristic of the problem is changed from compute-bound to memory-bound and high DRAM utilization (94\%) of 8 \acp{NPU} in Fig. \ref{fig:6_eval_multi_npu_latency} (b) proves this.

\noindent\textbf{Sensitivity to sequence length.}
Fig. \ref{fig:6_eval_input_sequence} illustrates the performance gain across varying input sequence lengths. Overall, we observe the similar trend on varying the input sequence length, however, when token length increases, the performance improvement from \textit{LUT} is more prominent (1K:18.1\% $\Rightarrow$ 4K:28.4\%) since the nonlinear portion increases quadratically.
For the same reason, the performance gain from \textit{Mask.opt} is also increased (1K:8.2\% $\Rightarrow$ 4K:8.9\%), however, FlashAttention \cite{dao2024flashattention2} scheduling, due to the large sequence length, diminishes the effect of \textit{Mask.opt} slightly.

\section{Conclusion} \label{sec:conclusion}
This paper presents TriGen \ac{NPU} architecture for end-to-end acceleration of \ac{LLM} inferences based on SW-HW co-design.
For efficient inferences, TriGen adopts MXINT8 number system and HW resource constraint aware scheduling. Also, the new \ac{LUT} architecture accelerates various nonlinear activation functions with negligible accuracy loss.
Various software optimizations such as QKV batching, fusing transpose operation and coalesced masking operation further improve the efficiency of TriGen inference.
The extensive experimental results demonstrate that TriGen achieves 2.73$\times$ speedup and reduces memory transfer by 52\% on average.


\bibliographystyle{IEEEtranS}
\bibliography{references}

\end{document}